______________________________________________________________________________
\documentstyle[12pt]{article}
\begin{document}

\title{Exact Solutions to the Schr\"{o}dinger Equation for the potential 
$V(r)=a r^2+b r^{-4}+c r^{-6}$ in 2D}

\author{Shi-Hai Dong\thanks{Electronic address: DONGSH@BEPC4. IHEP. AC. CN}\\
{\scriptsize Institute of High Energy Physics, P. O. Box 918(4), 
Beijing 100039, People's Republic of China}\\
\\
Zhong-Qi Ma\\
{\scriptsize China Center for Advanced Science and Technology
(World Laboratory), P. O. Box 8730, Beijing 100080}\\
{\scriptsize  and Institute of High Energy Physics, P. O. Box 918(4), 
Beijing 100039, People's Republic of China}}

\date{}

\maketitle

\begin{abstract}
Making use of an ${\it ansatz}$ for 
the eigenfunctions, we obtain an exact closed form
solution to the non-relativistic Schr\"{o}dinger 
equation with the anharmonic potential, 
$V(r)=a r^2+b r^{-4}+c r^{-6}$ in two dimensions, where 
the parameters of the potential $a, b, c$ satisfy some constraints. 

\end{abstract}

\vskip 6mm


\vskip 4mm


\newpage

\begin{center}
{\large 1. Introduction}\\
\end {center}

The exact solutions to the fundamental dynamical equations play crucial roles 
in physics. It is well-known that the exact solutions 
to the Schr\"{o}dinger equation 
are possible only for the several 
potentials and some approximation methods
are frequently applied to arrive 
at the solutions. On the other hand, 
in recent  years, the higher 
order anharmonic potentials have attracted 
much more attention to physicists and 
mathematicians[1-3]. Interest in these anharmonic oscillator-like
interactions stems from the fact that, in many case, the study of the
relevant Schr\"{o}dinger equation, for example in atomic and molecular
physics, provides us with insight into the physical problem in question.

Recall that in the three-dimensional spaces, rough speaking, 
there are two main methods to be used to 
deal with the anharmonic potentials 
$V(r)=a r^2+b r^{-4}+c r^{-6}$. One[4, 5] is 
based on an ${\it ansatz}$ for the
eigenfunctions to obtain an exact solution with this
potential. 
This method undoubtedly provides an exact
solution for the ground state but 
sometimes with some constraints on the
parameters of the potential. The other[6, 7] is 
relied on a Laurent series ${\it ansatz}$
for the eigenfunctions, which converts 
the Schr\"{o}dinger equation into a
difference equation and then the 
continued fraction solutions are defined. 
This method, however, does not 
give any constrains for the
parameters of the potential. 

The reasons why we write this paper are as follows. 
On the one hand, with the advent of growth 
technique for the realization of
the semiconductor quantum wells, the 
quantum mechanics of low-dimensional
systems has become a major research field. 
Almost all of the computational
technique developed for the three-dimensional 
problems has already been
extended to lower dimensions. On the other hand, 
the study of the potential 
$V(r)=a r^2+b r^{-4}+c r^{-6}$ in two 
dimensions has never been appeared
in the literature. We now attempt
to research it in two dimensions. 

This paper is organized as follows. 
In Sec. 2, we study the ground state
of the Schr\"{o}dinger equation with 
this potential using an  ${\it ansatz}$ for the eigenfunctions. 
The first excited state will be
discussed by the same way in Sec. 3. The some constraints
on the parameters of the potential $a, b, c$ are 
given in Secs. 2 and 3. The figures for the unnormalized radial functions 
are plotted in the last section.

\vskip 1cm
\begin{center}
{\large 2. The ground states }
\end{center}

Throughout this paper the natural unit $\hbar=1$ 
and $\mu=1/2$ are employed. 
Consider the Schr\"{o}dinger equation with a potential $V(r)$
that depends only on the distance $r$ from the origin
$$H\psi =\left(
\displaystyle {1 \over r} \displaystyle {\partial \over \partial r} 
r \displaystyle {\partial \over \partial r} + 
\displaystyle {1 \over r^{2}} \displaystyle {\partial^{2} \over 
\partial \varphi^{2} } \right)\psi +V(r) \psi =E \psi, \eqno(1)$$ 

\noindent
where the potential
$$V(r)=a r^2+b r^{-4}+c r^{-6}, ~~ a>0, ~~c>0. \eqno(2)$$

\noindent
Owing to the symmetry of the potential, let
$$\psi(r, \varphi)=r^{-1/2} R_{m}(r) e^{ \pm im \varphi}, 
~~~~~m=0, 1, 2, \ldots, \eqno (3) $$  

\noindent
where the radial wave function $R_{m}(r)$ satisfies the 
radial equation
$$\displaystyle {d^{2} R_{m}(r) \over dr^{2} }
+\left[E-V(r)-\displaystyle {m^{2}-1/4 \over r^{2}} \right] R_{m}(r)
=0, \eqno (4a) $$

\noindent
where $m$ and $E$ denote the 
angular momentum and energy, respectively. 
For the solution of Eq. (4a), we make an ${\it ansatz}$[4, 5] 
for the ground state
$$R_{m0}(r)=\exp[p_{m0}(r)], \eqno(5)$$
 
\noindent
where
$$p_{m0}(r)=\frac{1}{2} \alpha r^2+\frac{1}{2}\beta r^{-2}+\kappa \ln r. 
\eqno(6a)$$

\noindent
After calculating, we arrive at the following equation
$$\displaystyle {d^{2} R_{m0}(r) \over dr^{2}}
-\left[\displaystyle{d^{2} p_{m0}(r) \over dr^{2}}
+\left(\displaystyle {dp_{m0}(r) \over dr}\right)^2\right]R_{m0}(r)=0. 
\eqno(4b)$$

\noindent
Compare Eq. (4b) with Eq. (4a) and 
obtain the following set of equations
$$\alpha ^2=a, ~~~ \beta ^2=c, \eqno(7a)$$
$$\kappa ^2-\kappa-2\alpha \beta=m^2-1/4, \eqno(7b)$$
$$3\beta-2\beta \kappa=b, \eqno(7c)$$
$$E=-(2\kappa+1)\alpha. \eqno(7d)$$
\noindent
It is easy to obtain the values of parameters for $p_{m0}(r)$
from the Eqs. (7a) and (7b) written as
$$\alpha=\pm \sqrt{a}; ~~~~~ \beta=\pm \sqrt{c}; 
~~~~~\kappa=\frac{1}{2}\pm \sqrt{m^2+2\sqrt{ac}} . \eqno(8)$$

\noindent
In order to retain the well-behaved
solution at the origin and at infinity, 
we choose positive sign  in $\kappa$ and
negative signs  in $\alpha$
and $\beta$. According to these choices, 
the Eq. (7c) leads to the following 
constraint on the parameters of the potential, 
$$(b+2\sqrt{c})^2-4c(m^2+2\sqrt{ac})=0. \eqno(9)$$

\noindent
The eigenvalue $E$, however, will be given from Eq. (7d) as
$$E=\sqrt{a}\left(4+\displaystyle{\frac{b}{\sqrt{c}}}\right). \eqno(10)$$

\noindent
Now, the corresponding eigenfunctions Eq. (5) can be read as
$$R_{m0}(r)=N_{0}r^{\kappa}
exp\left[-\frac{1}{2}(\sqrt{a}r^2+\sqrt{c}r^{-2})\right], \eqno(11)$$

\noindent
where $N_{0}$ is the normalized constant. Here and hereafter 
$\kappa=1/2+\sqrt{m^2+2\sqrt{ac}}$. 

\vskip 1cm
\begin{center}
{\large 3. The first excited states }\\
\end{center}

With the same spirit, we make 
an ${\it ansatz}$ for the eigenfunctions corresponding
to the first excited state in the potential (2) as follows
$$R_{m1}(r)=f_{m}(r)\exp[p_{m1}(r)], \eqno(12)$$

\noindent
with $f_{m}(r)$ given by
$$f_{m}(r)=a_{1}+a_{2} r^2+a_{3} r^{-2}, \eqno(13)$$

\noindent
and $p_{m1}(r)$ given by
$$p_{m1}(r)=
\frac{1}{2} \alpha_{1} r^2+\frac{1}{2}\beta_{1} r^{-2}+\kappa_{1} \ln r. 
\eqno(6b)$$

\noindent
For short, it is readily to see
from Eqs. (12) and (13) that the radial function $R_{m1}(r)$ has
the following relation
$$R_{m1}(r)''-\left[p_{m1}(r)''+(p_{m1}(r)')^2
+\left(\frac{f_{m}(r)''+2p_{m1}(r)'f_{m}(r)'}{f_{m}(r)}\right)
\right]R_{m1}(r)=0, \eqno(4c)$$

\noindent
where the prime denotes the derivative
of the radial function with respect to
the variable $r$.
Calculating Eq. (4c) carefully
and comparing it with Eq. (4a), we obtain 
$$a_{2}[E-\sqrt{a}(2\kappa_{1}+5)]=0,~~~~ a_{3}[b-\sqrt{c}(2\kappa_{1}-7)]=0, 
\eqno(14a)$$
$$a_{1} [E-\sqrt{a}(2\kappa_{1}+1)]
=a_{2}[m^{2}-1/4+2\sqrt{ac}-\kappa_{1}^{2}-3\kappa_{1}-2], \eqno(14b)$$
$$a_{1}[m^{2}-1/4+2\sqrt{ac}-\kappa_{1}^{2}+\kappa_{1}]=
a_{2}[b-\sqrt{c}(2\kappa_{1}+1)]+a_{3}[-E+\sqrt{a}(2\kappa_{1}-3)], \eqno(14c)$$
$$ a_{1}[b-\sqrt{c}(2\kappa_{1}-3)]=
-a_{3}[m^{2}-1/4+2\sqrt{ac}-\kappa_{1}^{2}+5\kappa_{1}-6], \eqno(14d)$$
$$\alpha_{1}=\pm \sqrt{a}, ~~~~~~\beta_{1}=\pm \sqrt{c}. \eqno(14e)$$

\noindent
Hence, if the angular momentum $m$ of the first 
excited state is the same as that of
the ground state, we obtain from Eq. (14)
$$\kappa_{1}=\displaystyle{\frac{b+7\sqrt{c}}{2\sqrt{c}}}, 
~~~~E_{1}=\sqrt{a}(5+2\kappa_{1}), \eqno(15a)$$
$$a_{1}=0,~~~ a_{2}=\sqrt{a},~~~ a_{3}=-\sqrt{c}, \eqno(15b)$$
$$\beta_{1}=-\sqrt{c},~~~~~~~ \alpha_{1}=-\sqrt{a}, \eqno(15c)$$
$$b=-6\sqrt{c}, \eqno(15d)$$

\noindent
where the constants both $\alpha_{1}$ and $\beta_{1}$ are chosen negative
signs in order to hold well-behaved nature of the 
solution at $r\rightarrow 0$ and 
$r\rightarrow \infty$. 
Equation (15d) is another constrain 
on the parameters of the potential. 

At last, the eigenvalue $E_{1}$ and 
eigenfunctions $R_{m1}(r)$ for the first
excited state with the potential (2) may 
be read from Eqs. (15a) and (12) as follows
$$E_{1}=\sqrt{a}\left(12+\frac{b}{\sqrt{c}}\right), \eqno(16)$$

$$R_{m1}(r)=N_{1}(a_{2} r^2+a_{3} r^{-2})r^{\kappa_{1}}
\exp\left[-\frac{1}{2}(\sqrt{a}r^2+\sqrt{c}r^{-2})\right], \eqno(17)$$

\noindent
where $N_{1}$ is the normalized
constant for the first excited state 
and $\kappa_{1}$ is given by Eq. (15a). 

As a matter of fact, the normalized 
constants $N_{0}$ and $N_{1}$ can be
calculated in principle from the normalized relation
$$\int_{0}^{\infty}|R_{mi}|^2dr=1, ~~~~~ {\rm i}=0, 1. \eqno(18)$$

Considering the values of the
parameters of the potential, we fix them as follows. 
The value of parameter $a$ is first fixed, for example $a=1. 0$, 
the values of the parameter $c$ and $b$ are determined
by the constraints Eq. (9) and 
Eq. (15d) for $m=0$. By this way, 
the corresponding parameters turn out to
$a=1. 0, c=4, b=-12, \kappa=-1. 5, \kappa_{1}=0. 5, a_{2}=1, a_{3}=-2$. 
The ground state and the first excited state
energies corresponding to these
values are obtained as $E_{0}=-2$
and $E_{1}=6$, 
respectively. Actually, when we
study the properties of the ground state and the first
excited state, as we know, the unnormalized
radial wave functions will not
affect the main features of the wave functions. 
We have plotted the unnormalized
radial wave functions $R_{0}^{i}, (i=0, 1)$
in fig. 1 and fig. 2 for the ground 
state and the first excited states, respectively. 
Comparing them with the figures for the 
ground state and the first excited state in three dimensions, respectively,
it is easy to find that they are different from each other. The reason is that
the parameters of the potential 
$b, c$ are not same as those in three
dimensions, which origins from 
the different constraints on the parameters
of the potential, 
even if the parameter $a$ is same 
both in two dimensions and in
three dimensions. 

To summarize, we discuss the ground 
state and the first excited state
for the Schr\"{o}dinger 
equation with the potential 
$V(r)=a r^2+b r^{-4}+c r^{-6}$ using
a simpler ${\it ansatz}$ for 
the eigenfunctions and simultaneously 
two constrains for the
parameters of the potential are arrived at from the 
compared equations, which then results in
the variety for the energy eigenvalue 
and eigenfunctions with the varieties of
the parameters of the potential. 
This simple and intuitive method is 
easy to be generalized. The other studies to 
the sextic potential and the octic 
potential as well as the inverse 
potential in two dimensions are in progress.

\vspace{10mm}
{\bf Acknowledgments}. This work was supported by the National
Natural Science Foundation of China and Grant No. LWTZ-1298 from
the Chinese Academy of Sciences. 


\end{document}